\documentclass[12pt,twoside]{article}

\usepackage{epsf}
\usepackage{times}
\usepackage{a4wide}

\usepackage{fancyheadings,graphics}
\advance \headheight by 3.0truept       

\newlength{\nseparation}
\newenvironment{nfigure}[1]
        {\begin{figure}[#1]\hrule\vspace{\nseparation}\par}
        {\vspace{\nseparation}\par \hrule \end{figure}}
\newenvironment{ntable}[1]
        {\begin{table}[#1]\hrule\vspace{\nseparation}\par}
        {\vspace{\nseparation}\par \hrule \end{table}}

\usepackage{cite}   

\newcommand{\ds}{\displaystyle}
\newcommand{\lt}{\left}
\newcommand{\rt}{\right}
\newcommand{\no}{\nonumber}
\newcommand{\nn}{\nonumber \\}
\newcommand{\li}{\mathrm{Li}_2}
\newcommand{\su}{_}

\newcommand{\ov}[1]{\overline{#1}}
\newcommand{\eq}[1]{(\ref{#1})}
\newcommand{\imag}{\mathrm{Im}\,}

\newcommand{\gev}{\,\mbox{GeV}}
\newcommand{\mev}{\,\mbox{MeV}}
\newcommand{\WA}{u}
\newcommand{\PI}{d}

\begin{document}
\thispagestyle{empty}
\boldmath

\vspace*{-2cm}

\begin{flushright}
PITHA 02/05\\
CERN-TH/2002-019\\
BUTP-02/2\\
FERMILAB-Pub-02/016-T\\
hep-ph/0202106\\
February 2002 
\end{flushright}

\vspace*{1cm}

\centerline{\LARGE\bf The B$^+$--B$_{\rm d}^0$ Lifetime Difference} 
\vspace*{0.3cm}
\centerline{\LARGE\bf Beyond Leading Logarithms}
\unboldmath

\vspace*{1cm}
\centerline{\sc
   Martin Beneke$^1$,\, Gerhard Buchalla$^2$,\, Christoph~Greub$^3$,} 
\centerline{\sc   Alexander~Lenz$^4$\, and\,  Ulrich Nierste$^{2,5}$}

\vspace*{0.5cm}
\centerline{
\parbox{0.85\textwidth}{
\sl $^1$ Institut f\"ur Theoretische Physik E, RWTH Aachen,
    Sommerfeldstra\ss e 28,\\
    \phantom{$^1$} D-52074 Aachen, Germany.\\[2mm]
    $^2$Theory Division, CERN, CH-1211 Geneva 23,
                Switzerland.\\[2mm]
    $^3$Institut f\"ur Theoretische Physik, Universit\"at Bern,
     Sidlerstrasse 5,    \\
    \phantom{$^1$} CH-3012 Berne, Switzerland.\\[2mm]
    $^4$Fakult\"at f\"ur Physik, Universit\"at Regensburg, 
        D-93040 Regensburg, Germany.\\[2mm]
    $^5$Fermi National Accelerator Laboratory, Batavia, 
        IL 60510-500, USA. }}

\vspace*{1cm}
\centerline{\bf Abstract}
\vspace*{0.3cm}
\noindent 
We compute perturbative QCD corrections to the lifetime splitting
between the charged and neutral $B$ meson in the framework of
the heavy quark expansion. These next-to-leading logarithmic
corrections are necessary for a meaningful use of hadronic matrix
elements of local operators from lattice gauge theory. We find the
uncertainties associated with the choices of renormalization scale and
scheme significantly reduced compared to the leading-order result. 
We include the full dependence on the charm-quark mass $m_c$ without any
approximations.
Using hadronic matrix elements estimated in the
literature with lattice QCD we obtain 
$\tau(B^+)/\tau(B^0_d)=1.053\pm 0.016\pm 0.017$, where the effects of 
unquenching and $1/m_b$ corrections are not yet included.
The lifetime difference of heavy baryons $\Xi^0_b$ and $\Xi^-_b$
is also briefly discussed.

\vspace*{0.3cm}

\vfill

 
\newpage
\pagenumbering{arabic}
\setcounter{page}{2}

\section{Preliminaries}
The \emph{Heavy Quark Expansion}\ (HQE) technique provides a
well-defined QCD-based framework for the calculation of total decay
rates of $b$-flavoured hadrons \cite{hqe}. The HQE yields an expansion
of the decay rate $\Gamma(H_b)$ in terms of $\Lambda_{QCD}/m_b$, where
$H_b$ represents any hadron containing a single $b$-quark and any of
the light $u$,$d$,$s$ (anti-)quarks as valence quarks. $m_b$ is the
$b$-quark mass and $\Lambda_{QCD}$ is the fundamental scale of QCD,
which determines the size of hadronic effects. In the leading order of
$\Lambda_{QCD}/m_b$ the decay rate of $H_b$ equals the 
decay rate of a free $b$-quark, which is unaffected by the light
degrees of freedom of $H_b$. Consequently, the lifetimes of all
$b$-flavoured hadrons are the same at this order. The first
corrections to the free quark decay appear at order
$(\Lambda_{QCD}/m_b)^2$ and are caused by the Fermi motion of the 
$b$-quark in $H_b$ and the chromomagnetic interaction of the final state
quarks with the hadronic cloud surrounding the heavy $b$-quark. These
mechanisms have a negligible effect on the lifetime difference between
the $B^+$ and $B_d^0$ mesons, because the strong interaction
excellently respects isospin symmetry. At order
$(\Lambda_{QCD}/m_b)^3$, however, one encounters weak interaction
effects between the $b$-quark and the light valence quark. These
effects, known as \emph{weak annihilation}\ (WA) and
\emph{Pauli interference}\ (PI) \cite{hqe}, are depicted in
Fig.~\ref{fig:lo}.
\begin{nfigure}{t!}
\centerline{
\epsfxsize=0.8\textwidth
\epsffile{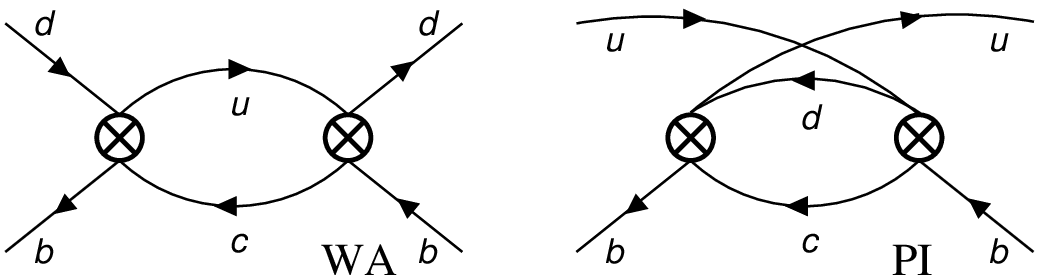}
}
\caption{\textit{Weak annihilation}\ (WA) 
  and \textit{Pauli interference}\ (PI) diagrams in the leading order
  of QCD. They contribute to $\Gamma (B_d^0)$ and $\Gamma (B^+)$,
  respectively.  The crosses represent $|\Delta B|\!=\!1$ operators,
  which are generated by the exchange of $W$ bosons. CKM-suppressed
  contributions are not shown.}\label{fig:lo}
\end{nfigure} 
They are phase-space enhanced with respect to the leading free-quark
decay and induce corrections to $\Gamma(H_b)$ of order $16 \pi^2
(\Lambda_{QCD}/m_b)^3 ={\cal O} (5\!\!-\!\!10\% )$.  The measurement
of lifetime differences among different $b$-flavoured hadrons
therefore tests the HQE formalism at the third order in the expansion
parameter.

The calculation of $\Gamma(H_b)$ consists of three steps: the first
step is an operator product expansion (OPE) integrating out  
the heavy $W$ boson, which mediates the weak $b$ decay.
This results in an effective $|\Delta B|=1$ Hamiltonian describing the
flavour-changing weak interaction of the Standard Model up to
corrections of order $m_b^2/M_W^2$,
where $\Delta B$ denotes the change in bottom-quark number:
\begin{eqnarray}
H &=& \frac{G_F}{\sqrt{2}} \, V_{cb}^* 
        \sum_{ {\scriptstyle d^\prime=d,s} \atop 
               {\scriptstyle u^\prime=u,c}
              } 
         V_{u^\prime d^\prime} \, 
         \lt[   C_1 (\mu_1) \, Q_1^{u^\prime d^\prime} (\mu_1) 
        \; + \; C_2 (\mu_1) \, Q_2^{u^\prime d^\prime} (\mu_1) 
          \rt]  \; + \; \mbox{h.c.}
    \label{heff} .
\end{eqnarray}
Here $G_F$ is the Fermi constant and the $V_{ij}$'s are elements of the
Cabibbo-Kobayashi-Maskawa (CKM) matrix. The Wilson coefficients $C_i(\mu_1)$ 
contain the short-distance physics associated with scales above the
renormalization scale $\mu_1$. The weak interaction is encoded in the
four-quark operators 
\begin{eqnarray}
Q_1^{u^\prime d^\prime} = \ov{b}_i \gamma_\mu (1-\gamma_5) c_j \,
                  \ov{u}^\prime_j \gamma^\mu (1-\gamma_5) d_i^\prime ,
 \quad && 
Q_2^{u^\prime d^\prime} = \ov{b}_i \gamma_\mu (1-\gamma_5) c_i \,
                  \ov{u}^\prime_j \gamma^\mu (1-\gamma_5) d_j^\prime,
\label{q1q2} 
\end{eqnarray} 
with summation over the colour indices $i$ and $j$. We have omitted
penguin operators and doubly Cabibbo-suppressed terms in \eq{heff},
which have a negligible effect on the $B^+$--$B_d^0$ lifetime
difference.  Next the total decay rate $\Gamma (H_b)$ is
related to $H$ by the optical theorem:
\begin{eqnarray}
\Gamma (H_b) &=& 
  \frac{1}{2 M_{H_b}} \langle H_b | {\cal T} | H_b \rangle .
  \label{opt}
\end{eqnarray}
Here we have adopted the conventional relativistic normalization 
$\langle H_b | H_b \rangle =2EV$ and introduced the transition operator:
\begin{eqnarray}
{\cal T} &=&  \imag i \! \int \!\! d^4 x \, 
              T [H(x) \, H(0)] 
                .  \label{deft}
\end{eqnarray}
The second step is the HQE, which exploits the hierarchy $m_b \gg
\Lambda_{QCD}$ to expand the RHS of \eq{opt} in terms of
$\Lambda_{QCD}/m_b$. To this end an OPE is applied to ${\cal T}$ which
effectively integrates out the hard loop momenta (corresponding to the
momenta of the final state quarks). 
We decompose the result as 
\begin{eqnarray}
{\cal T} &=& \lt[ {\cal T}_0 \; + \; {\cal T}_2 \; + \; {\cal T}_3 \rt] 
              \lt[ 1 \; + \; 
                      {\cal O} ( 1/m_b^4 )
              \rt] \nn 
{\cal T}_3 &=& {\cal T}^{\WA} + {\cal T}^{\PI} + {\cal T}_{sing}  
\label{t3}
\end{eqnarray}
Here ${\cal T}_n$ denotes the portion of ${\cal T}$ which is
suppressed by a factor of $1/m_b^n$ with respect to ${\cal T}_0$
describing the free quark decay. The contributions
to ${\cal T}_3$ from weak spectator interactions read
\begin{eqnarray}
{\cal T}^{\WA} &=& \frac{G_F^2 m_b^2 |V_{cb}|^2}{6 \pi} 
     \lt[\;\,\, |V_{ud}|^2 
     \lt(F^{\WA} Q^d \; + \; F_S^{\WA} Q_S^d \; + \;
          G^{\WA} T^d \; + \; G_S^{\WA} T_S^d \, \rt) \rt. \nn
&& \phantom{\frac{G_F^2 m_b^2 |V_{cb}|^2}{6 \pi} }
    \lt. +\, |V_{cd}|^2 
    \lt( \, F^c Q^d \,\, + \,\, F^c_S Q_S^d \; + \,\, G^c T^d \; + \;\,
 G^c_S T_S^d \, \rt)
     \rt] \, + \, (d \to s) \nn
{\cal T}^{\PI} &=& \frac{G_F^2 m_b^2 |V_{cb}|^2}{6 \pi} 
     \lt[ \,  F^{\PI} Q^u \; + \; F_S^{\PI} Q_S^u \; + \; 
          G^{\PI} T^u \; + \;  
          G_S^{\PI} T_S^u \, \rt] . \label{ope}  
\end{eqnarray}
The superscript $q$ of the coefficients $F^q$, $F^q_S$, $G^q$, $G^q_S$
refers to the $cq$ intermediate state (see Fig. \ref{fig:lo}).
We include singly Cabibbo-suppressed contributions.
In writing ${\cal T}^{\PI}$ we have used $|V_{ud}|^2+|V_{us}|^2\approx 1$
and $m_d\approx m_s\approx 0$, so that $F^d=F^s$, etc..
Here we encounter the local dimension-6, $\Delta B=0$ operators 
\begin{eqnarray}
Q^q   \; = \; \ov{b} \gamma_\mu (1-\gamma_5) q \, 
              \ov{q} \gamma^\mu (1-\gamma_5) b,~~~~~~~~&& \quad 
Q_S^q \; = \; \ov{b} (1-\gamma_5) q \, \ov{q} (1+\gamma_5) b, \no\\[1mm] 
T^q   \; = \;   \ov{b} \gamma_\mu (1-\gamma_5) T^a q \, 
       \ov{q} \gamma^\mu (1-\gamma_5) T^a b, && \quad
T_S^q \; = \; \ov{b} (1-\gamma_5) T^a q \, \ov{q} (1+\gamma_5) T^a b,
\label{ops}
\end{eqnarray}
where $T^a$ is the generator of colour SU(3). We define the $\Delta
B=0$ operators at the renormalization scale $\mu_0$, which is of order
$m_b$.  The Wilson coefficients $F^{\WA}\ldots G_S^{\PI}$ are computed
in perturbation theory. When applied to mesons, ${\cal T}^{\WA}$ and
${\cal T}^{\PI}$ correspond to the WA and PI mechanisms of
Fig.~\ref{fig:lo}, respectively. In the case of baryons their role is
interchanged: ${\cal T}^{\WA}$ encodes the PI effect and ${\cal
  T}^{\PI}$ describes the weak scattering of the $b$-quark with the
valence quark (see Fig.~\ref{fig:bar}).  The coefficients in \eq{ope}
depend on $\mu_0$.  Since the hard loops involve the charm quark, they
also depend on the ratio $z=m_c^2/m_b^2$.  The truncation of the
perturbation series makes $F^{\WA}\ldots G_S^{\PI}$ also dependent on
$\mu_1={\cal O}(m_b)$.  This dependence diminishes in increasing
orders of $\alpha_s$. To the considered order,
the dependence on $\mu_0$ cancels between the coefficients and 
the matrix elements of operators in \eq{ope}, so that observables
are independent of $\mu_0$.  The remainder ${\cal T}_{sing}$ 
in \eq{t3} involves additional dimension-6 operators, which
describe power-suppressed contributions to the free quark decay from
strong interactions with the spectator quark.  
The operators in ${\cal T}_{sing}$ are isospin singlets and do
not contribute to the $B^+$--$B_d^0$ lifetime difference.
The formalism of \eq{t3}--\eq{ops}
applies to weakly decaying hadrons containing a single bottom quark
and no charm quarks. Decays of hadrons like the $B_c$ meson with more
than one heavy quark have a different power counting than in \eq{t3}
\cite{bb}.  
In the third step one computes the hadronic matrix elements of the
operators in \eq{ops}. They enter our calculation in isospin-breaking
combinations and are conventionally parametrized as \cite{ns}
\begin{eqnarray}
\!\!\!\!\!\!\!\!\!&&
\langle B^+ | (Q^u - Q^d) (\mu_0) | B^+ \rangle \, = \,  
              f_B^2 M_B^2 B_1 (\mu_0),\;\; 
\langle B^+ | (Q_S^u - Q_S^d) (\mu_0) | B^+ \rangle \, = \,  
              f_B^2 M_B^2 B_2 (\mu_0), \nn 
\!\!\!\!\!\!\!\!\!&&\langle B^+ | (T^u - T^d) (\mu_0) | B^+ \rangle \, = \,  
              f_B^2 M_B^2 \epsilon_1 (\mu_0),\quad  
\langle B^+ | (T_S^u - T_S^d) (\mu_0) | B^+ \rangle \, = \,  
              f_B^2 M_B^2 \epsilon_2 (\mu_0). 
              \label{b} 
\end{eqnarray}
Here $f_B$ is the $B$ meson decay constant. In the \emph{vacuum
  saturation approximation}\ (VSA) one has $B_1(\mu_0)=1$,
$B_2(\mu_0)=1+{\cal O}(\alpha_s(m_b),\Lambda_{QCD}/m_b)$ and
$\epsilon_{1,2}(\mu_0)=0$.  Corrections to the VSA results are of
order $1/N_c$, where $N_c=3$ is the number of colours.

Using the isospin relation $\langle B_d^0 | Q^{d,u} | B_d^0 \rangle =
\langle B^+ | Q^{u,d} | B^+ \rangle $ we now find from \eq{opt} and
\eq{ope}:
\begin{eqnarray}
\Gamma (B_d^0) - \Gamma (B^+) & = & 
  \frac{G_F^2 m_b^2 |V_{cb}|^2}{12 \pi } \, f_B^2 M_B \, 
        \lt( |V_{ud}|^2 \vec{F}{}^{\WA} + |V_{cd}|^2 \vec{F}{}^{c} 
         - \vec{F}{}^{\PI} \rt) \cdot \vec{B} 
    . \label{diffg}
\end{eqnarray}
Here we have introduced the shorthand notation
\begin{eqnarray}
 \vec{F}{}^q (z, \mu_0 ) \;  = \; 
\lt(
 \begin{array}{c}
   F^q (z, \mu_0) \\
   F_S^q (z, \mu_0) \\
   G^q (z, \mu_0) \\
   G_S^q (z, \mu_0) 
 \end{array} 
\rt) , && \quad
\vec{B} (\mu_0 ) \;  = \; 
\lt(
\begin{array}{c}
  B_1 (\mu_0) \\
  B_2 (\mu_0) \\
  \epsilon_1 (\mu_0)\\
  \epsilon_2 (\mu_0)
\end{array}           
\rt) \qquad \mbox{for } q=\PI,\WA,c   . \label{short} 
\end{eqnarray}

The strong interaction affects all three steps of the calculation. The
minimal way to include QCD effects is the leading logarithmic
approximation, which includes corrections of order $\alpha_s^n \ln^n
(\mu_1/M_W)$, $n=0,1,\ldots$ in the coefficients
$C_{1,2}(\mu_1)$ in \eq{heff}. The corresponding leading order (LO)
calculation of the width difference in \eq{diffg} involves the
diagrams in Fig.~\ref{fig:lo} \cite{hqe,ns}.  Yet LO results are too
crude for a precise calculation of lifetime differences. The heavy-quark
masses in \eq{diffg} cannot be defined in a proper way and one faces a
large dependence on the renormalization scale $\mu_1$.
Furthermore, results for $B_{1,2}$ and $\epsilon_{1,2}$ from lattice
gauge theory cannot be matched to the continuum theory in a meaningful
way at LO. Finally, as pointed out in \cite{ns}, at LO the
coefficients $F$, $F_S$ in \eq{diffg} are anomalously small. They
multiply the large matrix elements parametrized by $B_{1,2}$, while
the larger coefficients $G$, $G_S$ come with the small hadronic
parameters $\epsilon_{1,2}$, rendering the LO prediction highly
unstable.  To cure these problems one must include the
next-to-leading-order (NLO) QCD corrections of order $\alpha_s^{n+1}
\ln^n (\mu_1/M_W)$.  NLO corrections to the effective $|\Delta B|=1$
Hamiltonian in \eq{heff} have been computed in \cite{acmp,bw}.  The
second step beyond the LO requires the calculation of QCD corrections
to the coefficients $F^{\WA}\ldots G_S^{\PI}$ in \eq{ope}. Such a
calculation has been first performed for the $B_s^0$--$B_d^0$ lifetime
difference in \cite{kn}, where ${\cal O}(\alpha_s)$ corrections were
calculated in the SU(3)$_{\rm F}$ limit neglecting certain terms of order
$z$. In this limit only a few penguin effects play a role. A complete
NLO computation has been carried out for the lifetime difference
between the two mass eigenstates of the $B_s^0$ meson in \cite{bbgln}.
In particular the correct treatment of infrared effects, which appear
at intermediate steps of the calculation, has been worked out in
\cite{bbgln}.  The computation presented in this paper is conceptually
similar to the one in \cite{bbgln}, except that the considered
transition is $\Delta B=0$ rather than $\Delta B=2$ and the quark
masses in the final state are different. 
While this work was in preparation, QCD corrections to ${\cal
  T}^{\WA}$ and ${\cal T}^{ \PI}$ have also been calculated 
in \cite{rome}. There are two important differences between our
analysis and \cite{rome}:
\begin{itemize} 
\item[(i)] in \cite{rome} the NLO corrections have been computed for
  the limiting case $z=0$, i.e.\ neglecting the charm-quark mass in
  the final state. The corrections to this limit are of order $z\ln z$
  or roughly 20\%.  In Sect.~\ref{sect:nlo} we include the 
  dependence on the charm-quark mass exactly.
\item[(ii)] in \cite{rome} the $\Delta B=0$ operators have been
  defined in the heavy quark effective theory (HQET) rather than in
  full QCD, as we did in \eq{ops}. HQET operators were chosen to
  eliminate the mixing of the dimension-6 operators in \eq{ops} into 
  lower-dimensional operators under renormalization. We emphasize that
  this mixing does not impede the use of QCD operators in the HQE:
  it results purely from ultraviolet effects and can be accounted
  for by a finite renormalization of the affected operators. For 
  a more detailed discussion with an explicit example we refer the
  reader to \cite{bbgln} and to Sect.~\ref{subs:xi}.   
\end{itemize} 
Finally one must compute the non-perturbative QCD effects residing in
$f_B^2 B_1,\ldots f_B^2\epsilon_2$. Results from lattice
gauge theory for the matrix elements in \eq{b} have been recently
obtained in \cite{b}. Earlier results using HQET fields can be found
in \cite{ds}. In the matching of the results to continuum QCD the
dependence of $B_1,\ldots \epsilon_2$ on $\mu_0$ and on the chosen
renormalization scheme must cancel the corresponding dependence 
of the Wilson coefficients, which requires NLO accuracy.

\boldmath
\section{${\cal T}^{\WA}$ and ${\cal T}^{\PI}$ at
  next-to-leading order}\label{sect:nlo}
\unboldmath
We decompose the Wilson coefficients in \eq{ope} as  
\begin{eqnarray}
F^{\WA} (z,\mu_0) &=& \phantom{\; + \;}
          C_1^2 (\mu_1) \, F^{\WA}_{11} (z,x_{\mu_1},x_{\mu_0}) \; + \; 
    C_1 (\mu_1) \, C_2 (\mu_1)\,F^{\WA}_{12} (z,x_{\mu_1},x_{\mu_0}) \nn 
&& \; + \; 
          C_2^2 (\mu_1)\,F^{\WA}_{22} (z,x_{\mu_1},x_{\mu_0}) \no \\[1mm]
F^{\WA}_{ij} (z,x_{\mu_1},x_{\mu_0}) &=& F^{\WA,(0)}_{ij} (z) \; + \: 
          \frac{\alpha_s (\mu_1)}{4\pi} \, F^{\WA,(1)}_{ij}
          (z,x_{\mu_1},x_{\mu_0}) + {\cal O}\lt( \alpha_s^2 \rt) \label{ffij} 
\end{eqnarray}
with $x_{\mu}=\mu/m_b$ and an analogous notation for the remaining Wilson
coefficients in \eq{ope}.  The LO coefficients are obtained from the
diagrams in Fig.~\ref{fig:lo}. The non-vanishing coefficients read \cite{ns}
\begin{eqnarray}
&& \frac{1}{3} F_{11}^{\WA,(0)}(z) =  
\frac{1}{2} F_{12}^{\WA,(0)}(z) =  
3 F_{22}^{\WA,(0)}(z) =  
\frac{1}{2} G_{22}^{\WA,(0)}(z) =  
- \left( 1 - z \right)^2\,\left( 1 + \frac{z}{2}\right) , \nonumber\\
&& \frac{1}{3} F_{S,11}^{\WA,(0)}(z) =  
\frac{1}{2} F_{S,12}^{\WA,(0)}(z) =  
3 F_{S,22}^{\WA,(0)}(z) =  
\frac{1}{2} G_{S,22}^{\WA,(0)}(z) =  
\left( 1 - z \right)^2\, \left( 1 + 2 z \right) , \nonumber\\
&&\frac{1}{3} F_{11}^{c,(0)}(z) =  
\frac{1}{2} F_{12}^{c,(0)}(z) =  
3 F_{22}^{c,(0)}(z) =  
\frac{1}{2} G_{22}^{c,(0)}(z) =  
-\sqrt{1- 4z} \, \left( 1 - z \right) , \\
&& \frac{1}{3} F_{S,11}^{c,(0)}(z) =  
\frac{1}{2} F_{S,12}^{c,(0)}(z) =  
3 F_{S,22}^{c,(0)}(z) =  
\frac{1}{2} G_{S,22}^{c,(0)}(z)=  
\sqrt{1- 4z} \, \left( 1 + 2 z \right) , \nonumber\\[0.1cm]
&& 6 F_{11}^{\PI,(0)} (z) = F_{12}^{\PI,(0)} (z) = 6 F_{22}^{\PI,(0)} (z) = 
G_{11}^{\PI,(0)} (z) = G_{22}^{\PI,(0)} (z) 
=  6 \left( 1 - z \right)^2,\nonumber
\end{eqnarray} 
while
\begin{eqnarray}
&& G_{11}^{\WA,(0)} =  G_{12}^{\WA,(0)} =  
G_{S,11}^{\WA,(0)} =  G_{S,12}^{\WA,(0)} = G_{11}^{c,(0)} =  G_{12}^{c,(0)} =  
G_{S,11}^{c,(0)} =  G_{S,12}^{c,(0)} = G_{12}^{\PI,(0)} =  0, \nonumber\\
&& F_{S,ij}^{\PI,(0)} =  
   G_{S,ij}^{\PI,(0)} = 0.
\end{eqnarray}

To obtain the NLO corrections $F^{\WA,(1)}_{ij} \ldots G_{S,ij}^{
  \PI,(1)}$ we have calculated the diagrams $E_i$ and the
imaginary parts of $D_i$ in Fig.~\ref{fig:nlo}.
\begin{nfigure}{t!}
\centerline{
\epsfxsize=\textwidth
\epsffile{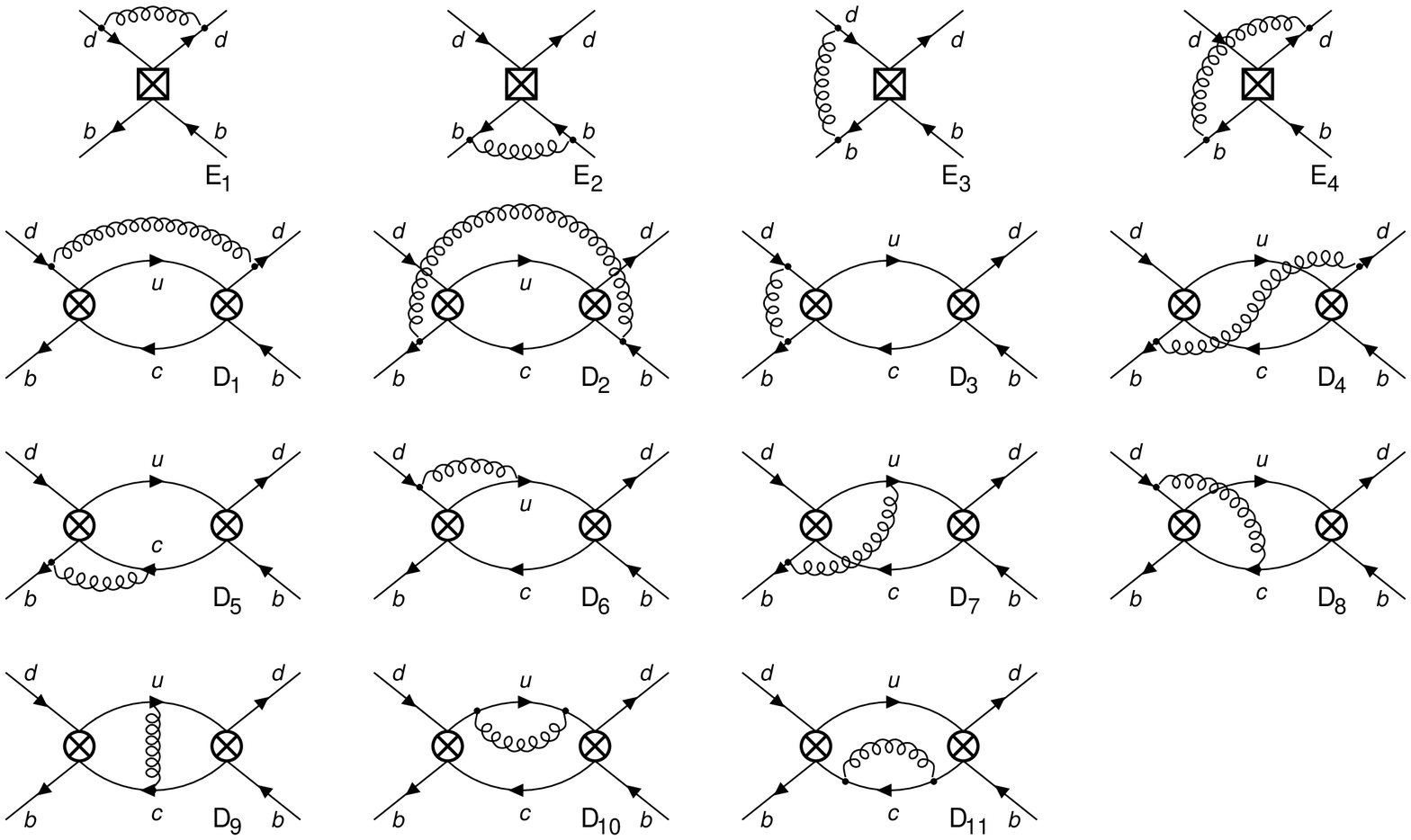}
}
\caption{WA contributions in the next-to-leading order of QCD. The PI
diagrams are obtained by interchanging $u$ and $d$ and reversing the 
fermion flow of the $u$ and $d$ lines. The first line shows the 
radiative corrections to $\Delta B\!=\!0$ operators, which are
necessary for the proper infrared factorization. 
Not displayed are the diagrams ${\rm E}_3^\prime$, ${\rm E}_4^\prime$
and ${\rm D}_{3-8}^\prime$ which are obtained from the corresponding 
unprimed diagrams by left-right reflection and the reverse of the
fermion flow.}\label{fig:nlo}
\end{nfigure}
At NLO one becomes sensitive to the renormalization scheme.  First,
this affects the quantities $m_b$, $z$ and $\alpha_s$
entering our calculation. The NLO coefficients given below correspond
to the use of the pole-mass definition for $m_b$ and the
definition of $\alpha_s$ in the $\ov{\rm MS}$ scheme \cite{bbdm}.  $z$
can be either calculated from the pole masses or from the $\ov{\rm
  MS}$ masses, because $z=m_c^2/m_b^2=\ov{m}_c^2(m_c)/\ov{m}_b^2 (m_b)
+ {\cal O} (\alpha_s^2) $. Second, the choice of the
renormalization scheme is also an issue for the effective four-quark
operators appearing at the various stages of our calculation. In the
prediction of physical quantities this scheme dependence cancels to
the calculated order, nevertheless it must be taken care of when
assembling pieces from different theoretical sources. The Wilson
coefficients $C_{1,2}$ of $H$ in \eq{heff} and
$F^{\WA,(1)}_{ij} \ldots G_{S,ij}^{\PI,(1)}$ depend on the scheme used
to renormalize the $\Delta B=1$ operators in \eq{q1q2}, but this
dependence cancels in $F^{\WA,(1)} \ldots G_S^{\PI,(1)}$. Our results
below correspond to the definition of $C_{1,2}$ in \cite{bw}. 
$F^{\WA,(1)} \ldots G_S^{\PI,(1)}$ also depend on the
renormalization scheme of the $\Delta B=0$ operators in \eq{ops}. This
dependence cancels only when these coefficients are combined with the
hadronic parameters $B_{1,2}$ and $\epsilon_{1,2}$ calculated from
lattice QCD. It is therefore important that our scheme is used in the
lattice-continuum matching of these quantities. We use the $\ov{\rm
  MS}$ scheme with the NDR prescription for $\gamma_5$ \cite{bw}. 
To specify the
scheme completely, it is further necessary to state the definition of
the evanescent operators appearing in the calculation \cite{hn}. We
use
\begin{eqnarray}
E[Q]   &=& \ov{b} \gamma_\mu \gamma_\rho \gamma_\nu (1-\gamma_5) q \, 
         \ov{q} \gamma^\nu \gamma^\rho \gamma^\mu (1-\gamma_5) b 
         \; - \; 
         (4 - 8 \varepsilon) \, Q \nn
E[Q_S] &=& \ov{b} \gamma_\mu \gamma_\nu (1-\gamma_5) q \, 
         \ov{q} \gamma^\nu \gamma^\mu (1+\gamma_5) b \; - \;  
         (4 - 8 \varepsilon) \, Q_S \label{defev}
\end{eqnarray}
and analogous definitions of $E[T]$ and $E[T_S]$. When the diagrams
${\rm E}_1\ldots {\rm E}_4$ for e.g.\ $Q_S$ are calculated in $D=4-2
\varepsilon$ dimensions, the result can be expressed as a linear
combination of $Q_S$ and $E[Q_S]$. 
Effectively, \eq{defev} defines how Dirac strings with
two or three Dirac matrices are reduced. (Note that \eq{defev} also
implies the replacement rules $\gamma_\nu \gamma_\rho \gamma_\mu
(1-\gamma_5) \otimes \gamma^\nu \gamma^\rho \gamma^\mu (1-\gamma_5)
\to (16 - 4 \varepsilon ) \gamma_\mu (1-\gamma_5) \otimes \gamma^\mu
(1-\gamma_5)$ and $ \gamma_\mu \gamma_\nu (1-\gamma_5) \otimes
\gamma^\mu \gamma^\nu (1+\gamma_5) \to 4 (1+\varepsilon) (1-\gamma_5)
\otimes (1+\gamma_5)$.) The particular choice of the $- 8\varepsilon $
terms in \eq{defev} is motivated by Fierz invariance: the
one-loop matrix elements of e.g.\ $Q_S$ and its Fierz transform
$Q_S^F=-1/2\, \ov{b}_i\gamma_\nu (1+\gamma_5) b_j \ov{q}_j \gamma^\nu
(1-\gamma_5) q_i$ are in general different. This feature is an artifact of
dimensional regularization. With \eq{defev} and a corresponding
definition of $E[Q_S^F]$, however, Fierz invariance is maintained at
the one-loop level. This choice, which has also been made in \cite{bw}
for the $\Delta B=1$ operators, has the practical advantage that
one can freely use the Fierz transformation at any step of the
calculation. In other words: ``Fierz-evanescent'' operators like
$Q_S-Q_S^F$ can be identified with 0.
 
In the procedure of matching the full theory (eq. (\ref{deft})) to the
effective $\Delta B=0$ theory, infrared singularities are
encountered at ${\cal O}(\alpha_s)$ both in the full-theory diagrams
and in the matrix elements of operators in the effective theory.
The diagrams relevant for this issue are $D_1$ -- $D_4$ and
$E_1$ -- $E_4$. The singularities cancel in the Wilson coefficients
$F$ and $G$, but need to be regularized at intermediate steps of the
calculation. We take the $b$-quark on-shell, assign zero $4$-momentum
to the external light quarks and use dimensional regularization for
the infrared (as well as the ultraviolet) divergences. In this case,
care has to be taken to treat the Dirac algebra in a consistent way.
In computing the matching condition between  $D_1$ -- $D_4$ and
$E_1$ -- $E_4$ we have used two different methods, which lead
to the same result.
In both methods ultraviolet divergences appearing in  $E_1$ -- $E_4$
and $D_3$ are subtracted, respectively, by $\Delta B=0$ and
$\Delta B=1$ counterterms, in the usual way.

In the first method, we distinguish IR singularities arising in loop
integrals from UV singularities, and treat the Dirac algebra in strictly
four dimensions in the IR-divergent parts.
In the second method, IR and UV divergences are not distinguished
and $d$-dimensional Dirac algebra is used throughout. In this case
evanescent operators $E$, as those given in (\ref{defev}), give a 
non-vanishing contribution in the matching procedure. This is a
subtlety of the IR regulator used in method 2 \cite{MU}. If a different
IR regulator, such as a gluon mass or method 1, is used, the
non-vanishing bare one-loop matrix element of $E$ is cancelled by a
finite counterterm, so that $E$ disappears from the NLO matching
calculation \cite{bw,hn}. The non-zero contribution in method 2
originates in diagram $E_1$ with the insertion of an evanescent
operator $E$. This diagram is zero in dimensional regularization, 
thus leaving the corresponding counterterm uncancelled.
We have further parametrized the evanescent ${\cal O}(\varepsilon)$
parts appearing in the $d$-dimensional projections of general Dirac
structures $\Gamma\otimes\Gamma$ onto the basic operators $Q$ and $Q_S$.
There are four independent parameters in the calculation, corresponding
to $\Gamma$ being a string of two, three, four or five Dirac matrices.
We have checked that all four parameters disappear from the final
result for the coefficients. (This is true for the evanescent
${\cal O}(\varepsilon)$ parts multiplying IR poles.
The UV poles give rise to a dependence on these parameters, which
corresponds to a usual scheme dependence that is cancelled by the
matrix elements of operators in the effective theory. Our choice
of scheme is specified by (\ref{defev}).)

We would also like to mention that the Fierz ordering of
$\Delta B=1$ operators is immaterial because Fierz symmetry
is respected by the standard NDR renormalization scheme employed
by us. This has been checked by using the Fierz form leading to
Dirac strings with flavour structure $\bar bb\otimes \bar uu$
in method 1, and $\bar bu\otimes\bar ub$ in method 2, and
similarly for the contribution with $u\to d$. (The Fierz form
used in method 2 for $\bar bd\otimes\bar db$ is such that a
closed fermion loop is generated in $D_1$ -- $D_4$.)

In the NLO corrections to \eq{diffg} we set $|V_{ud}|=1$ and
$V_{cd}=0$. This introduces an error of order $|V_{cd}|^2
\alpha_s(m_b) z \ln z$, which is well below 1\% of 
$\tau(B^+)/\tau(B^0_d)-1$. Hence \eq{diffg} only involves the
differences $F^{\WA,(1)}_{ij} - F^{\PI,(1)}_{ij} \ldots
G^{\WA,(1)}_{S,ij} - G^{\PI ,(1)}_{S,ij}$.  Our results for these
coefficients read:
\begin{eqnarray}
\lefteqn{
F^{\WA ,(1)}_{11} (z,x_{\mu_1},x_{\mu_0}) - 
  F^{\PI ,(1)}_{11} (z,x_{\mu_1},x_{\mu_0})
 = } \no \\[2mm] 
 &&
\ds \left[
\frac{16\,\left( 1 - z \right) \,\left( -4 - 3\,z + 3\,z^2 \right) }{3}
 \right] \; \ds \left[\li(z) + \frac{\ln (1 - z)\,\ln (z)}{2}\right] \; + \no \\[2mm]
 &&\ds \left[
\frac{4\,{\left( 1 - z \right) }^2\,\left( 16 + 19\,z \right) }{3}
 \right] \; \ds \ln (1 - z) \; + \;
\ds \left[
\frac{4\,z\,\left( 93 + 40\,z - 57\,z^2 \right) }{9}
 \right] \; \ds \ln (z) \; + \no \\[2mm]
 &&\ds \left[
32\,{\left( 1 - z \right) }^2
 \right] \; \ds \ln (x\su {\mu\su 1}) \; + \;
\ds \left[
-16\,{\left( 1 - z \right) }^2
 \right] \; \ds \ln (x\su {\mu\su 0}) \; + \no \\[2mm]
 &&\ds \left[
\frac{32\,\left( 1 - z \right) }{9}
 \right] \; \ds {\pi }^2 \; + \;
\ds
\frac{2\,\left( 1 - z \right) \,\left( 152 + 149\,z + 155\,z^2 \right) }{27}
\no 
\end{eqnarray}

\begin{eqnarray}
\lefteqn{
F^{\WA ,(1)}_{12} (z,x_{\mu_1},x_{\mu_0}) - 
  F^{\PI ,(1)}_{12} (z,x_{\mu_1},x_{\mu_0})
 = } \no \\[2mm] 
 &&
\ds \left[
\frac{32\,\left( 1 - z \right) \,\left( -4 - 6\,z + z^2 \right) }{3}
 \right] \; \ds \left[\li(z) + \frac{\ln (1 - z)\,\ln (z)}{2}\right] \; + \no \\[2mm]
 &&\ds \left[
\frac{8\,{\left( 1 - z \right) }^2\,\left( 2 + 13\,z + 3\,z^2 \right) }{3\,z}
 \right] \; \ds \ln (1 - z) \; + \;
\ds \left[
\frac{8\,z\,\left( 37 - 6\,z - 6\,z^2 \right) }{3}
 \right] \; \ds \ln (z) \; + \no \\[2mm]
 &&\ds \left[
16\,{\left( 1 - z \right) }^2\,\left( 2 + z \right) 
 \right] \; \ds \ln (x\su {\mu\su 1}) \; + \;
\ds \left[
\frac{16\,\left( 1 - z \right) \,\left( 6 + 2\,z + z^2 \right) }{9}
 \right] \; \ds {\pi }^2 \; + \no \\[2mm]
 &&\ds
\frac{4\,\left( 1 - z \right) \,\left( 30 + 33\,z - 13\,z^2 \right) }{3}
\no 
\end{eqnarray}

\begin{eqnarray}
\lefteqn{
F^{\WA ,(1)}_{22} (z,x_{\mu_1},x_{\mu_0}) - 
  F^{\PI ,(1)}_{22} (z,x_{\mu_1},x_{\mu_0})
 = } \no \\[2mm] 
 &&
\ds \left[
\frac{16\,\left( 19 - z \right) \,\left( -1 + z \right) \,z}{9}
 \right] \; \ds \left[\li(z) + \frac{\ln (1 - z)\,\ln (z)}{2}\right] \; + \no \\[2mm]
 &&\ds \left[
\frac{16\,{\left( 1 - z \right) }^2\,{\left( 1 + 2\,z \right) }^2}{9\,z}
 \right] \; \ds \ln (1 - z) \; + \;
\ds \left[
\frac{4\,z\,\left( 135 + 30\,z - 68\,z^2 \right) }{27}
 \right] \; \ds \ln (z) \; + \no \\[2mm]
 &&\ds \left[
\frac{16\,{\left( 1 - z \right) }^2\,\left( 8 + z \right) }{3}
 \right] \; \ds \ln (x\su {\mu\su 1}) \; + \;
\ds \left[
\frac{-8\,{\left( 1 - z \right) }^2\,\left( 8 + z \right) }{3}
 \right] \; \ds \ln (x\su {\mu\su 0}) \; + \no \\[2mm]
 &&\ds \left[
\frac{16\,\left( 1 - z \right) \,\left( 6 + 2\,z + z^2 \right) }{27}
 \right] \; \ds {\pi }^2 \; + \;
\ds
\frac{4\,\left( 1 - z \right) \,\left( 544 - 185\,z - 68\,z^2 \right) }{81}
\no 
\end{eqnarray}

\begin{eqnarray}
\lefteqn{
F^{\WA ,(1)}_{S,11} (z,x_{\mu_1},x_{\mu_0}) - 
  F^{\PI ,(1)}_{S,11} (z,x_{\mu_1},x_{\mu_0})
 = } \no \\[2mm] 
 &&
\ds \left[
32\,{\left( 1 - z \right) }^2\,\left( 1 + 2\,z \right) 
 \right] \; \ds \left[\li(z) + \frac{\ln (1 - z)\,\ln (z)}{2}\right] \; + \no \\[2mm]
 &&\ds \left[
-8\,{\left( 1 - z \right) }^2\,\left( 2 + 10\,z - 3\,z^2 \right) 
 \right] \; \ds \ln (1 - z) \; + \no \\[2mm]
 &&\ds \left[
\frac{8\,z\,\left( 18 - 155\,z + 144\,z^2 - 27\,z^3 \right) }{9}
 \right] \; \ds \ln (z) \; + \no \\[2mm]
 &&\ds \left[
-48\,{\left( 1 - z \right) }^2\,\left( 1 + 2\,z \right) 
 \right] \; \ds \ln (x\su {\mu\su 0}) \; + \;
\ds
\frac{-4\,\left( 1 - z \right) \,\left( 133 - 53\,z + 40\,z^2 \right) }{27}
\no 
\end{eqnarray}

\begin{eqnarray}
\lefteqn{
F^{\WA ,(1)}_{S,12} (z,x_{\mu_1},x_{\mu_0}) - 
  F^{\PI ,(1)}_{S,12} (z,x_{\mu_1},x_{\mu_0})
 = } \no \\[2mm] 
 &&
\ds \left[
\frac{64\,\left( 1 - z \right) \,\left( 2 - z \right) \,
    \left( 1 + 2\,z \right) }{3}
 \right] \; \ds \left[\li(z) + \frac{\ln (1 - z)\,\ln (z)}{2}\right] \; + \no \\[2mm]
 &&\ds \left[
\frac{-16\,{\left( 1 - z \right) }^2\,
    \left( 1 + 2\,z + 6\,z^2 - 3\,z^3 \right) }{3\,z}
 \right] \; \ds \ln (1 - z) \; + \no \\[2mm]
 &&\ds \left[
\frac{16\,z\,\left( 4 - 24\,z + 18\,z^2 - 3\,z^3 \right) }{3}
 \right] \; \ds \ln (z) \; + \no \\[2mm]
 &&\ds \left[
-32\,{\left( 1 - z \right) }^2\,\left( 1 + 2\,z \right) 
 \right] \; \ds \ln (x\su {\mu\su 1}) \; + \;
\ds \left[
-32\,{\left( 1 - z \right) }^2\,\left( 1 + 2\,z \right) 
 \right] \; \ds \ln (x\su {\mu\su 0}) \; + \no \\[2mm]
 &&\ds \left[
\frac{-32\,\left( 1 - z \right) \,z\,\left( 1 + 2\,z \right) }{9}
 \right] \; \ds {\pi }^2 \; + \;
\ds
\frac{8\,\left( 1 - z \right) \,\left( -17 - 29\,z + 36\,z^2 \right) }{3}
\no 
\end{eqnarray}

\begin{eqnarray}
\lefteqn{
F^{\WA ,(1)}_{S,22} (z,x_{\mu_1},x_{\mu_0}) - 
  F^{\PI ,(1)}_{S,22} (z,x_{\mu_1},x_{\mu_0})
 = } \no \\[2mm] 
 &&
\ds \left[
\frac{32\,\left( 1 - z \right) \,\left( 3 - z \right) \,
    \left( 1 + 2\,z \right) }{9}
 \right] \; \ds \left[\li(z) + \frac{\ln (1 - z)\,\ln (z)}{2}\right] \; + \no \\[2mm]
 &&\ds \left[
\frac{-8\,{\left( 1 - z \right) }^2\,
    \left( 2 + 5\,z + 8\,z^2 - 3\,z^3 \right) }{9\,z}
 \right] \; \ds \ln (1 - z) \; + \no \\[2mm]
 &&\ds \left[
\frac{8\,z\,\left( 18 - 123\,z + 82\,z^2 - 9\,z^3 \right) }{27}
 \right] \; \ds \ln (z) \; + \no \\[2mm]
 &&\ds \left[
\frac{-32\,{\left( 1 - z \right) }^2\,\left( 1 + 2\,z \right) }{3}
 \right] \; \ds \ln (x\su {\mu\su 1}) \; + \;
\ds \left[
\frac{-16\,{\left( 1 - z \right) }^2\,\left( 1 + 2\,z \right) }{3}
 \right] \; \ds \ln (x\su {\mu\su 0}) \; + \no \\[2mm]
 &&\ds \left[
\frac{-32\,\left( 1 - z \right) \,z\,\left( 1 + 2\,z \right) }{27}
 \right] \; \ds {\pi }^2 \; + \;
\ds
\frac{4\,\left( 1 - z \right) \,\left( -259 - 421\,z + 488\,z^2 \right) }{81}
\no 
\end{eqnarray}

\begin{eqnarray}
\lefteqn{
G^{\WA ,(1)}_{11} (z,x_{\mu_1},x_{\mu_0}) - 
  G^{\PI ,(1)}_{11} (z,x_{\mu_1},x_{\mu_0})
 = } \no \\[2mm] 
 &&
\ds \left[
16\,\left( 4 - 3\,z \right) \,\left( 1 - z \right) 
 \right] \; \ds \left[\li(z) + \frac{\ln (1 - z)\,\ln (z)}{2}\right] \; + \no \\[2mm]
 &&\ds \left[
{\left( 1 - z \right) }^2\,\left( 122 + 5\,z \right) 
 \right] \; \ds \ln (1 - z) \; + \;
\ds \left[
\frac{z\,\left( 384 - 256\,z - 21\,z^2 \right) }{3}
 \right] \; \ds \ln (z) \; + \no \\[2mm]
 &&\ds \left[
-24\,{\left( 1 - z \right) }^2
 \right] \; \ds \ln (x\su {\mu\su 1}) \; + \;
\ds \left[
-6\,{\left( 1 - z \right) }^2\,\left( 4 + 3\,z \right) 
 \right] \; \ds \ln (x\su {\mu\su 0}) \; + \no \\[2mm]
 &&\ds \left[
\frac{4\,\left( 7 - 9\,z \right) \,\left( 1 - z \right) }{3}
 \right] \; \ds {\pi }^2 \; + \;
\ds
\frac{\left( 1 - z \right) \,\left( -2450 + 2575\,z + 517\,z^2 \right) }{18}
\no 
\end{eqnarray}

\begin{eqnarray}
\lefteqn{
G^{\WA ,(1)}_{12} (z,x_{\mu_1},x_{\mu_0}) - 
  G^{\PI ,(1)}_{12} (z,x_{\mu_1},x_{\mu_0})
 = } \no \\[2mm] 
 &&
\ds \left[
8\,\left( 4 - 13\,z \right) \,\left( 1 - z \right) 
 \right] \; \ds \left[\li(z) + \frac{\ln (1 - z)\,\ln (z)}{2}\right] \; + \no \\[2mm]
 &&\ds \left[
\frac{2\,{\left( 1 - z \right) }^2\,\left( 2 + 3\,z + 13\,z^2 \right) }{z}
 \right] \; \ds \ln (1 - z) \; + \;
\ds \left[
\frac{4\,z\,\left( 12 + 24\,z - 25\,z^2 \right) }{3}
 \right] \; \ds \ln (z) \; + \no \\[2mm]
 &&\ds \left[
12\,{\left( 1 - z \right) }^2\,\left( 14 + z \right) 
 \right] \; \ds \ln (x\su {\mu\su 1}) \; + \;
\ds \left[
-12\,{\left( 1 - z \right) }^2\,\left( 8 + z \right) 
 \right] \; \ds \ln (x\su {\mu\su 0}) \; + \no \\[2mm]
 &&\ds \left[
\frac{4\,\left( 1 - z \right) \,\left( 6 + 2\,z + z^2 \right) }{3}
 \right] \; \ds {\pi }^2 \; + \;
\ds
\frac{\left( 1 - z \right) \,\left( 818 - 667\,z - 19\,z^2 \right) }{9}
\no 
\end{eqnarray}

\begin{eqnarray}
\lefteqn{
G^{\WA ,(1)}_{22} (z,x_{\mu_1},x_{\mu_0}) - 
  G^{\PI ,(1)}_{22} (z,x_{\mu_1},x_{\mu_0})
 = } \no \\[2mm] 
 &&
\ds \left[
\frac{-8\,\left( 1 - z \right) \,\left( 36 + 31\,z + 5\,z^2 \right) }{3}
 \right] \; \ds \left[\li(z) + \frac{\ln (1 - z)\,\ln (z)}{2}\right] \; + \no \\[2mm]
 &&\ds \left[
\frac{4\,{\left( 1 - z \right) }^2\,\left( -1 + 68\,z + 5\,z^2 \right) }
  {3\,z}
 \right] \; \ds \ln (1 - z) \; + \;
\ds \left[
\frac{4\,z\,\left( 162 - 102\,z - z^2 \right) }{9}
 \right] \; \ds \ln (z) \; + \no \\[2mm]
 &&\ds \left[
-4\,{\left( 1 - z \right) }^2\,\left( 8 + z \right) 
 \right] \; \ds \ln (x\su {\mu\su 1}) \; + \;
\ds \left[
2\,{\left( 1 - z \right) }^2\,\left( 8 + z \right) 
 \right] \; \ds \ln (x\su {\mu\su 0}) \; + \no \\[2mm]
 &&\ds \left[
\frac{2\,\left( 1 - z \right) \,\left( 60 + 77\,z + 7\,z^2 \right) }{9}
 \right] \; \ds {\pi }^2 \; + \;
\ds
\frac{\left( 1 - z \right) \,\left( -2803 + 2786\,z + 725\,z^2 \right) }{27}
\no 
\end{eqnarray}

\begin{eqnarray}
\lefteqn{
G^{\WA ,(1)}_{S,11} (z,x_{\mu_1},x_{\mu_0}) - 
  G^{\PI ,(1)}_{S,11} (z,x_{\mu_1},x_{\mu_0})
 = } \no \\[2mm] 
 &&
\ds \left[
-18\,{\left( 1 - z \right) }^2\,\left( 1 + 2\,z \right) 
 \right] \; \ds \ln (1 - z) \; + \;
\ds \left[
\frac{-44\,\left( 4 - 3\,z \right) \,z^2}{3}
 \right] \; \ds \ln (z) \; + \no \\[2mm]
 &&\ds
\frac{4\,\left( 1 - z \right) \,\left( 28 + 103\,z - 164\,z^2 \right) }{9}
\no 
\end{eqnarray}

\begin{eqnarray}
\lefteqn{
G^{\WA ,(1)}_{S,12} (z,x_{\mu_1},x_{\mu_0}) - 
  G^{\PI ,(1)}_{S,12} (z,x_{\mu_1},x_{\mu_0})
 = } \no \\[2mm] 
 &&
\ds \left[
16\,\left( 1 - z \right) \,\left( 1 + 2\,z \right) 
 \right] \; \ds \left[\li(z) + \frac{\ln (1 - z)\,\ln (z)}{2}\right] \; + \no \\[2mm]
 &&\ds \left[
\frac{-4\,{\left( 1 - z \right) }^2\,\left( 1 + z \right) \,
    \left( 1 + 2\,z \right) }{z}
 \right] \; \ds \ln (1 - z) \; + \;
\ds \left[
\frac{4\,z\,\left( 6 - 51\,z + 28\,z^2 \right) }{3}
 \right] \; \ds \ln (z) \; + \no \\[2mm]
 &&\ds \left[
-24\,{\left( 1 - z \right) }^2\,\left( 1 + 2\,z \right) 
 \right] \; \ds \ln (x\su {\mu\su 1}) \; + \;
\ds \left[
\frac{-8\,\left( 1 - z \right) \,z\,\left( 1 + 2\,z \right) }{3}
 \right] \; \ds {\pi }^2 \; + \no \\[2mm]
 &&\ds
\frac{4\,\left( 1 - z \right) \,\left( -53 - 80\,z + 82\,z^2 \right) }{9}
\no 
\end{eqnarray}

\begin{eqnarray}
\lefteqn{
G^{\WA ,(1)}_{S,22} (z,x_{\mu_1},x_{\mu_0}) - 
  G^{\PI ,(1)}_{S,22} (z,x_{\mu_1},x_{\mu_0})
 = } \no \\[2mm] 
 &&
\ds \left[
\frac{16\,\left( 1 - z \right) \,\left( 1 + 2\,z \right) \,
    \left( 3 + 5\,z \right) }{3}
 \right] \; \ds \left[\li(z) + \frac{\ln (1 - z)\,\ln (z)}{2}\right] \; + \no \\[2mm]
 &&\ds \left[
\frac{-2\,{\left( 1 - z \right) }^2\,
    \left( -2 + 31\,z + 64\,z^2 + 3\,z^3 \right) }{3\,z}
 \right] \; \ds \ln (1 - z) \; + \no \\[2mm]
 &&\ds \left[
\frac{2\,z\,\left( 36 - 336\,z + 62\,z^2 + 9\,z^3 \right) }{9}
 \right] \; \ds \ln (z) \; + \no \\[2mm]
 &&\ds \left[
8\,{\left( 1 - z \right) }^2\,\left( 1 + 2\,z \right) 
 \right] \; \ds \ln (x\su {\mu\su 1}) \; + \;
\ds \left[
4\,{\left( 1 - z \right) }^2\,\left( 1 + 2\,z \right) 
 \right] \; \ds \ln (x\su {\mu\su 0}) \; + \no \\[2mm]
 &&\ds \left[
\frac{-4\,\left( 1 - z \right) \,\left( 1 + 2\,z \right) \,
    \left( 9 + 7\,z \right) }{9}
 \right] \; \ds {\pi }^2 \; + \;
\ds
\frac{\left( 1 - z \right) \,\left( 385 + 1519\,z - 3278\,z^2 \right) }{27}
\label{diffcoeff} 
\end{eqnarray}
Here $\li (z) = - \int_0^z \,dt \, [\ln (1-t)]/t$ is the dilogarithm
function.  
Any dependence on infrared regulators has cancelled from the
coefficients in \eq{diffcoeff} showing that infrared effects properly
factorize. As another check we have verified that the dependence on
$\mu_1$ cancels analytically to the calculated order.

For our numerical studies we choose the following range for the input
parameters: 
\begin{eqnarray}
\alpha_s (M_Z) &=& 0.118 \pm 0.003, \qquad m_b \; = \; 4.8 \pm 0.1
     \gev, \qquad z=0.085 \pm 0.015 . \label{inp}
\end{eqnarray}
Throughout this paper we always remove ${\cal
  O}(\alpha_s^2)$ terms from the calculated coefficients. (For instance, 
  at NLO we write a product such as $C_1^2 F^u$ as 
  $C_{1,\rm LO}^2 F^u_{\rm NLO}+2 C_{1,\rm LO} \,dC_{1} 
  F^u_{\rm LO}$, where $C_{1,\rm NLO}=C_{1,\rm LO}+dC_{1}$ 
denotes the NLO Wilson coefficient.)
In all terms we use the two-loop expression for the running
coupling $\alpha_s$ in QCD with
five flavours. Numerical
values for the calculated coefficients can be found in Tab.~\ref{tab}.
\begin{ntable}{tb}
\begin{displaymath}
\begin{array}{r|r|r|r|r|r}
\ds z\, & \multicolumn{3}{c|}{\ds 0.085} &\ds 0.070  &\ds 0.100 \\ 
\ds \mu_1\, &\ds  m_b/2 &\ds  m_b &\ds 2\, m_b          &\ds  m_b &\ds  m_b 
\\[1mm]\hline\hline &&&&&\\[-2mm] 
\ds F^{\WA ,{\rm LO}} - F^{\PI ,{\rm LO}}
&\ds   0.865&\ds   0.270&\ds  -0.176&\ds   0.280&\ds   0.261\\ 
\ds F^{\WA ,{\rm NLO}} - F^{\PI ,{\rm NLO}}
&\ds   0.396&\ds   0.460&\ds   0.386&\ds   0.469&\ds   0.452\\[1mm]\hline &&&&&\\
[-2mm] 
\ds F_S^{\WA ,{\rm LO}} - F_S^{\PI ,{\rm LO}}
&\ds   0.002&\ds   0.042&\ds   0.105&\ds   0.043&\ds   0.042\\ 
\ds F_S^{\WA ,{\rm NLO}} - F_S^{\PI ,{\rm NLO}}
&\ds   0.035&\ds   0.033&\ds   0.026&\ds   0.031&\ds   0.035\\[1mm]\hline &&&&&\\
[-2mm] 
\ds G^{\WA ,{\rm LO}} - G^{\PI ,{\rm LO}}
&\ds  -9.912&\ds  -8.618&\ds  -7.848&\ds  -8.887&\ds  -8.353\\ 
\ds G^{\WA ,{\rm NLO}} - G^{\PI ,{\rm NLO}}
&\ds  -8.665&\ds  -8.501&\ds  -8.154&\ds  -8.718&\ds  -8.280\\[1mm]\hline &&&&&\\
[-2mm] 
\ds G_S^{\WA ,{\rm LO}} - G_S^{\PI ,{\rm LO}}
&\ds   2.679&\ds   2.404&\ds   2.231&\ds   2.420&\ds   2.385\\ 
\ds G_S^{\WA ,{\rm NLO}} - G_S^{\PI ,{\rm NLO}}
&\ds   1.668&\ds   1.850&\ds   1.902&\ds   1.854&\ds   1.843

\end{array}
\end{displaymath}
\caption{Numerical values for the coefficients in \eq{diffg} for
$\alpha_s(M_Z)=0.118$ and $\mu_0=m_b=4.8\gev$.  }\label{tab}
\end{ntable}
The two contributions from $(F^{\WA } - F^{\PI})B_1 + (G^{\WA } - G^{\PI})
\epsilon_1$ and from $(F_S^{\WA } - F_S^{\PI})B_2 + (G_S^{\WA } - G_S^{\PI})
\epsilon_2$ to $\Gamma (B_d^0) - \Gamma (B^+)$ are separately
scheme-independent. Tab.~\ref{tab} reveals that the former part is
expected to give the dominant contribution to the desired width
difference.  The results also show a substantial improvement of the 
$\mu_1$-dependence in the NLO compared to LO. 
\begin{nfigure}{t!}
\centerline{
\epsfxsize=0.5\textwidth
\epsffile{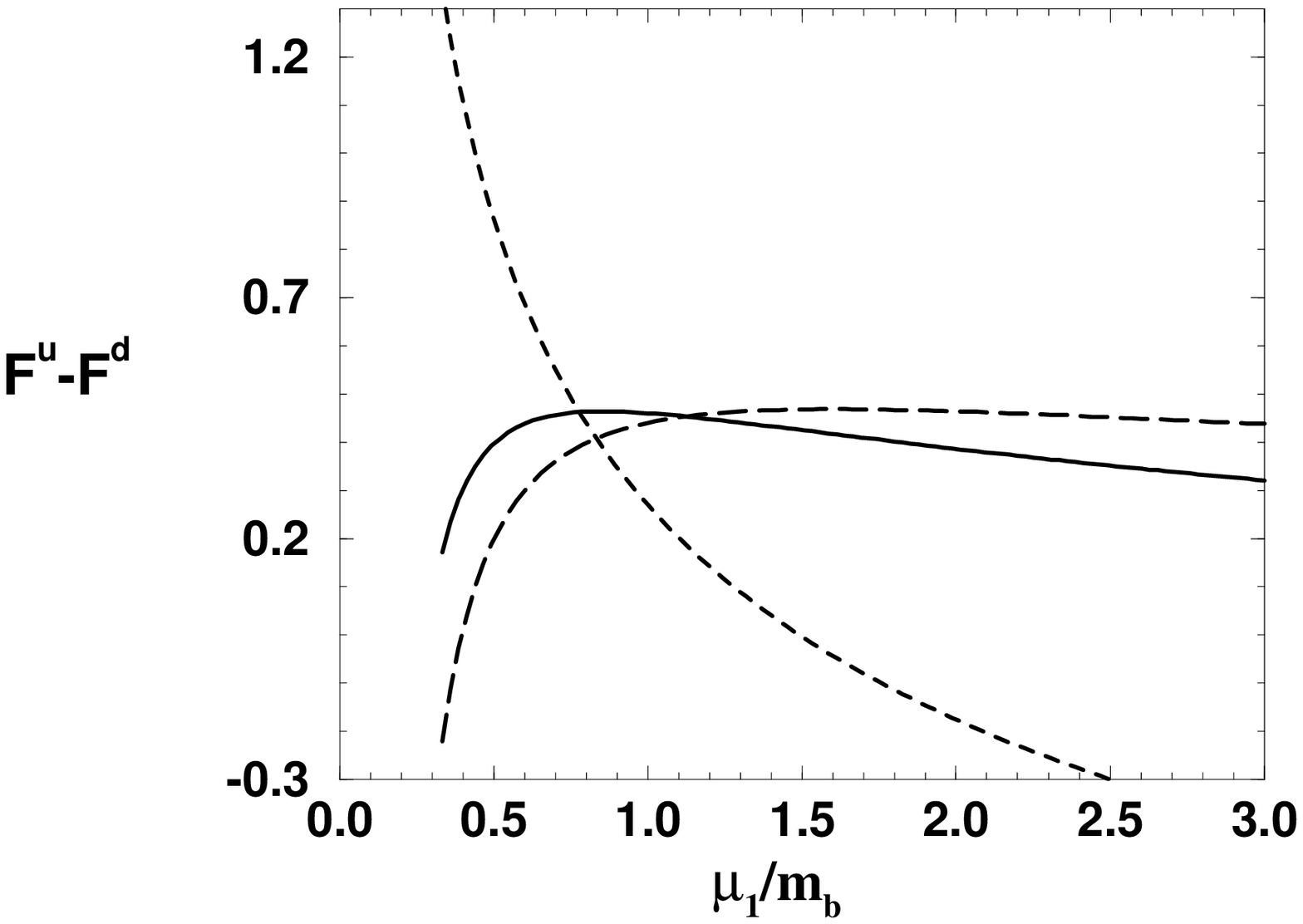}~~~
\epsfxsize=0.5\textwidth
\epsffile{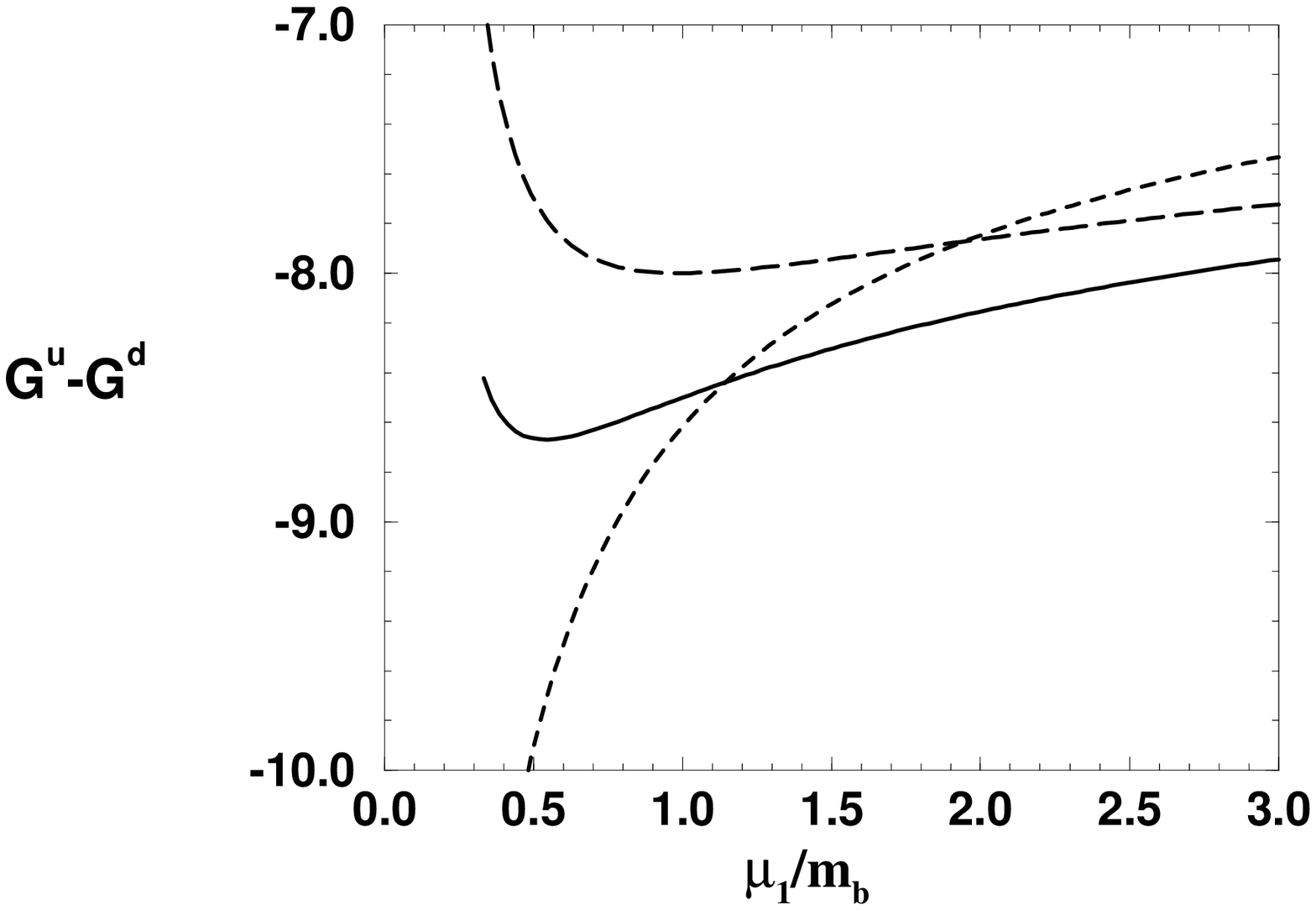}
}
\caption{Dependence of $ F^{\WA } - F^{\PI } $ and $ G^{\WA}
  - G^{\PI} $ on $\mu_1/m_b$ for the input parameters in \eq{inp} and
  $\mu_0=m_b$.  The solid (short-dashed) line shows the NLO (LO)
  result. The long-dashed line shows the NLO result in the
  approximation of \cite{rome}, i.e.\ $z$ is set to zero in the NLO
  corrections.  }\label{fig:plot}
\end{nfigure}
This dependence is plotted in Fig.~\ref{fig:plot} for the two Wilson
coefficients of the important vector operators.  The approximation
employed in \cite{rome} setting $z=0$ in the NLO correction is also
plotted. Expectedly, the accuracy of this approximation decreases for
small $\mu_1$, because the difference to the exact NLO result is of
order $\alpha_s(\mu_1)\, z\ln z$. For the final result of our
coefficients we estimate the $\mu_1$-dependence in a more conservative
way: we vary $\mu_1$ in $F^{\WA }\ldots G_S^{\WA }$ and $F^{\PI
  }\ldots G_S^{\PI }$ independently. Further the variation with $z$
and $\alpha_s(M_Z)$ in the ranges of \eq{inp} is calculated and all
these sources of theoretical uncertainty are symmetrized
individually and added in quadrature. The
dependence on $z$ is only an issue for $ G^{\WA} - G^{\PI} $. We find:
\begin{eqnarray}
\begin{array}{r|@{~~~}r|@{~~~}r|@{~~~}r}
& {\rm NLO}\phantom{0.10} & {\rm LO} \phantom{0.106}&
  {\rm app}\phantom{0.10} \\\hline &&&\\[-2mm]
F^{\WA } - F^{\PI } & 0.460 \pm 0.101 & 
                      0.270 \pm 0.480 &
                      0.440 \pm 0.119
\\[1mm]
F_S^{\WA } - F_S^{\PI } & 0.033 \pm 0.046 & 
                          0.042 \pm 0.052 &
                          0.025 \pm 0.045
\\[1mm]
G^{\WA } - G^{\PI } & -8.50 \pm 0.40~~   
                    & -8.62 \pm 0.90~~
                    & -8.00 \pm 0.32~~
\\[1mm]
G_S^{\WA } - G_S^{\PI } & 1.85 \pm 0.08~~
                        & 2.40 \pm 0.23~~  
                        & 1.80 \pm 0.10~~
\end{array}
\label{tabres}                     
\end{eqnarray}
The quoted central values correspond to the choice $\mu_1=m_b$ and the
central values in \eq{inp}. The third column in \eq{tabres} shows the
result for the approximation of \cite{rome}, setting $z=0$ in the NLO
corrections.  For $\mu_1=m_b$ this approximation reproduces the size
of the NLO corrections to $ F^{\WA } - F^{\PI } $ and $ G_S^{\WA} -
G_S^{\PI} $ to better than 15\% . The small NLO correction to $
G^{\WA } - G^{\PI } $ is, however, overestimated. The NLO result for
this coefficient, which is largest in magnitude, is better reproduced
by the LO result than by the approximation of \cite{rome}.

The origin of the $\alpha_s(\mu_1)\, z\ln z$ terms, which are the main
cause of the discrepancy between the first and third column in
\eq{tabres}, can be traced back to diagram ${\rm D}_{11}$ of
Fig.~\ref{fig:nlo}. This diagram defines the scheme of the charm-quark
mass. One can absorb the $\alpha_s(\mu_1)\, z\ln z$ terms into the LO
by replacing $z$ with $\ov{z}= \ov{m}^2_c(\mu)/\ov{m}^2_b(\mu) $,
which implies the replacement
\begin{eqnarray}
F^{\WA, (1)} \to  F^{\WA, (1)} - \frac{\alpha_s }{4\pi }
    \frac{\partial F^{\WA, (0)} }{\partial \ov{z}} 
    \, \gamma_m^{(0)} \, \ov{z} \ln \ov{z}  \label{sumlnz}
\end{eqnarray}
in the NLO corrections to $F^{\WA }$ and similarly in the other Wilson
coefficients. Here $\gamma_m^{(0)}=8$ is the LO anomalous dimension of
the quark mass. This procedure sums the terms of order $\alpha_s^n
(\mu_1)\, z\ln^n z$ with $n=0,1,\ldots$ to all orders in perturbation
theory. This can be seen by performing an OPE of the transition
operator ${\cal T}$ which treats $m_c$ as a light mass scale: then
increasing powers of $m_c$ correspond to $\Delta B=0$ operators of
increasing dimension and $m_c$ and $m_b$ enter the result at the same
scale $\mu_1$ at which the OPE is performed. In every order of the
perturbation series $\ln \ov{z}$ is split into $\ln (\mu_1^2/m_b^2)$
contained in the Wilson coefficients and $\ln (m_c^2/\mu_1^2)$
residing in the matrix elements.  Since there are no dimension-8
operators with charm-quark fields contributing to $\Gamma (B_d^0) -
\Gamma (B^+)$, no terms of order $m_c^2 \ln (m_c^2/\mu_1^2)$ can
occur.  From our NLO results we can indeed verify that the procedure
in \eq{sumlnz} removes the $\alpha_s(\mu_1)\, z\ln z$ terms, while
e.g.\ terms of order $\alpha_s(\mu_1)\, z^2 \ln z$ persist as
expected, because there are dimension-10 operators with charm-quark
fields of the type $m_c(\bar bq)(\bar qb)(\bar cc)$. 
Using $\ov{z}=0.055$ rather than $z=0.085$ in the coefficients
tabulated in the third column of \eq{tabres} indeed removes the
disturbing discrepancy with the NLO result for $G^{\WA } - G^{\PI }$.
Also the central values
of $F^{\WA } - F^{\PI }$ and $G_S^{\WA } - G_S^{\PI }$ move closer to the
NLO result, while no significant improvement occurs for $F_S^{\WA } -
F_S^{\PI }$.

The width difference in \eq{diffg} involves the product
$\vec{F}{}^{q\,T}\, \vec{B}$, which is independent of the
renormalization scheme and scales.  
In order to compare the scheme dependent coefficients
$\vec F^q$ with the calculation in \cite{rome}
for $z=0$, we need to take into account that the coefficients
in \cite{rome} are defined for matrix elements in HQET rather than
in full QCD. The matching relation connecting HQET and full-QCD
matrix elements of the four operators $\vec O$ used in \cite{rome}
has the form
\begin{equation}\label{oqcdhqet}
\langle\vec O\rangle_{QCD}(m_b)=
\left(1+\frac{\alpha_s(m_b)}{4\pi}\hat C^{\overline{MS}}_1\right)
\, \langle\vec O\rangle_{HQET}(m_b)\, ,
\end{equation}
where the $4\times 4$ matrix $\hat C^{\overline{MS}}_1$ can be found in
Eq. (36) of \cite{rome}. The renormalization
scheme of operator matrix elements in full QCD is identical
in our paper and in \cite{rome,b}. The only further difference
is that the operators $\vec O$ are linear combinations,
$\vec O=S\vec Q$, of our basis $\vec Q=(Q,\, Q_S,\, T,\, T_S)^T$
with
\begin{equation}\label{s44}
S=\left(
\begin{array}{cccc}
\frac{1}{3} & 0 & 2 & 0 \\
0 & -\frac{2}{3} & 0 & -4 \\
\frac{4}{9} & 0 & -\frac{1}{3} & 0 \\
0 & -\frac{8}{9} & 0 & \frac{2}{3} \\
\end{array}\right)\, .
\end{equation}
(This simple relation holds beyond tree level because the
renormalization schemes are identical. The preservation of
Fierz-symmetry by the choice of evanescent operators in (\ref{defev}) 
is important for this property.)
It follows that our coefficients $\vec F$ are related to the
corresponding coefficients $\vec A+\frac{\alpha_s}{4\pi}\vec B$
in \cite{rome} at scale $\mu=m_b$ through
\begin{equation}\label{fabrome}
\frac{1}{3}\left(\vec F^{(0)}+\frac{\alpha_s}{4\pi}\vec F^{(1)}\right)^T
=\vec A^{\, T}\, S+ \frac{\alpha_s}{4\pi}
\left(\vec B^T\, S - \vec A^{\,T}\, \hat C^{\overline{MS}}_1\, S\right)\, .
\end{equation}
Here we have suppressed flavour labels $q=u$, $d$ and the double
indices $ij=11$, $12$, $22$ refering to the $\Delta B=1$ coefficients
$C_i C_j$ (see (\ref{ffij})). Note that in the notation of
\cite{rome} labels $u$, $d$ are interchanged with respect to our
convention and that the coefficients with label $12$ are defined
with a relative factor of two.
Using (\ref{fabrome}) we have verified that the results
of \cite{rome} obtained for $z=0$ are in agreement with ours
in this limit.

\section{Phenomenology}
\boldmath
\subsection{$\tau(B^+)/\tau(B_d^0)$}
\unboldmath
One can directly use \eq{diffg} to predict the desired lifetime ratio:
\begin{eqnarray} \lefteqn{
  \frac{\tau(B^+)}{\tau(B_d^0)} - 1 \; = \;  \tau(B^+) \,
                       \lt[ \Gamma (B_d^0) - \Gamma (B^+) \rt]} \nn & = &
0.0325 \, \lt( \frac{|V_{cb}|}{0.04} \rt)^2
        \, \lt( \frac{m_b}{4.8\gev} \rt)^2 \,
           \lt( \frac{f_B}{200\mev} \rt)^2 \,
        \times \nn &&
      \Big[ \, ( 1.0 \pm 0.2) \, B_1 \; + \; (0.1 \pm 0.1) \, B_2 \; - \;
         (18.4 \pm 0.9) \, \epsilon_1 \; + \; (4.0 \pm 0.2) \, \epsilon_2
      \, \Big] .~~~ \label{phent}
\end{eqnarray}
Here $\tau(B^+) = 1.653\,$ps${}$ has been used in the overall
factor and the hadronic parameters $B_1\ldots \epsilon_2$ are
normalized at $\mu_0=m_b$ throughout this section. 

In \cite{ns} it has been noticed that without a detailed study of the 
hadronic parameters one expects $\tau(B^+)/\tau(B_d^0)$ to deviate
from 1 by up to $\pm 20\%$. This feature originates from the large
coefficient of $\epsilon_1$ and persists in our NLO prediction in 
\eq{phent}, because the NLO corrections to $G^{\WA } - G^{\PI }$
are small. Confronting \eq{phent} with the recent measurements 
 \cite{babar,belle},
\begin{eqnarray}
\frac{\tau(B^+)}{\tau(B_d^0)} &=& 
\left\{ \begin{array}{ll} 1.082 \pm 0.026 \pm 0.012 & {\rm (BABAR)} \\ 
                          1.091 \pm 0.023 \pm 0.014 & {\rm (BELLE)}
        \end{array}\right.
     \label{babar}
\end{eqnarray}
one expects $|\epsilon_1|$ to be significantly smaller than
$1/N_c=1/3$, i.e.\ nonfactorizable contributions appear to be small.
This result is confirmed by the existing computations of the
$\epsilon_i$'s in quenched lattice QCD \cite{b,ds}. 
However, due to its large coefficient 
sophisticated non-perturbative methods are
definitely necessary to compute $\epsilon_1$ sufficiently accurately.
The other important term in \eq{phent} is the first one: the NLO
enhancement of $F^{\WA} -F^{\PI}$ in \eq{tabres} has altered the
coefficient of $B_1$ in \eq{phent} from $0.6 \pm 1.0$ in the LO to
$1.0 \pm 0.2$. While from the LO result not even the sign of this
contribution was known, the NLO result now clearly establishes a positive
contribution of order 3\% to $\tau(B^+)/\tau(B_d^0)$ from the term
involving $B_1$.

The hadronic parameters have been computed in \cite{b} using
the same renormalization scheme as in the present paper.
They read 
\begin{eqnarray}
(B_1,B_2, \epsilon_1,\epsilon_2) &=& 
 (1.10\pm 0.20, \, 0.79 \pm 0.10, \, -0.02 \pm 0.02, \, 0.03 \pm
 0.01 ) . 
\label{bec}
\end{eqnarray}
Using $|V_{cb}|=0.040 \pm 0.0016 $ from a CLEO analysis of inclusive
semileptonic $B$ decays \cite{cleo}, the world average $f_B = (200 \pm
30)\mev $ from lattice calculations \cite{r} and $m_b=4.8 \pm 0.1
\gev$ in \eq{phent}, we find   
\begin{eqnarray}
\frac{\tau(B^+)}{\tau(B_d^0)} &=& 1.053 \pm 0.016 \pm 0.017, \qquad
\lt[ \frac{\tau(B^+)}{\tau(B_d^0)} \rt]_{\rm LO} \; =\; 1.041 \pm 0.040 
\pm 0.013,
\label{res}
\end{eqnarray}
where the first error is due to the errors on the NLO coefficients 
as given in (\ref{phent}) and the hadronic parameters (\ref{bec}),  
and the second error is 
the overall normalization uncertainty due to $m_b$, $|V_{cb}|$ and 
$f_B$ in (\ref{phent}). The first error reduces to 0.008 in NLO 
and 0.038 in LO, if the errors on the hadronic parameters 
are neglected, demonstrating the substantial reduction of
scale dependence at NLO in comparison with the LO. 
This result is gratifying as the strong scale 
dependence observed at LO had been a major motivation for a 
NLO analysis. This is also seen in Fig. \ref{fig:ldplot}, where  
we show the lifetime ratio as a function
of the renormalization scale $\mu_1$. We should, however, 
emphasize that the result and error given in (\ref{res}) do 
not include the effects of $1/m_b$ corrections and unquenching, 
which could well be on the order of 0.05. 
The NLO result slightly exceeds the central value of the LO result and
improves the agreement with the experimental value in \eq{babar}.

\begin{nfigure}{t!}
\centerline{
\epsfxsize=0.7\textwidth
\epsffile{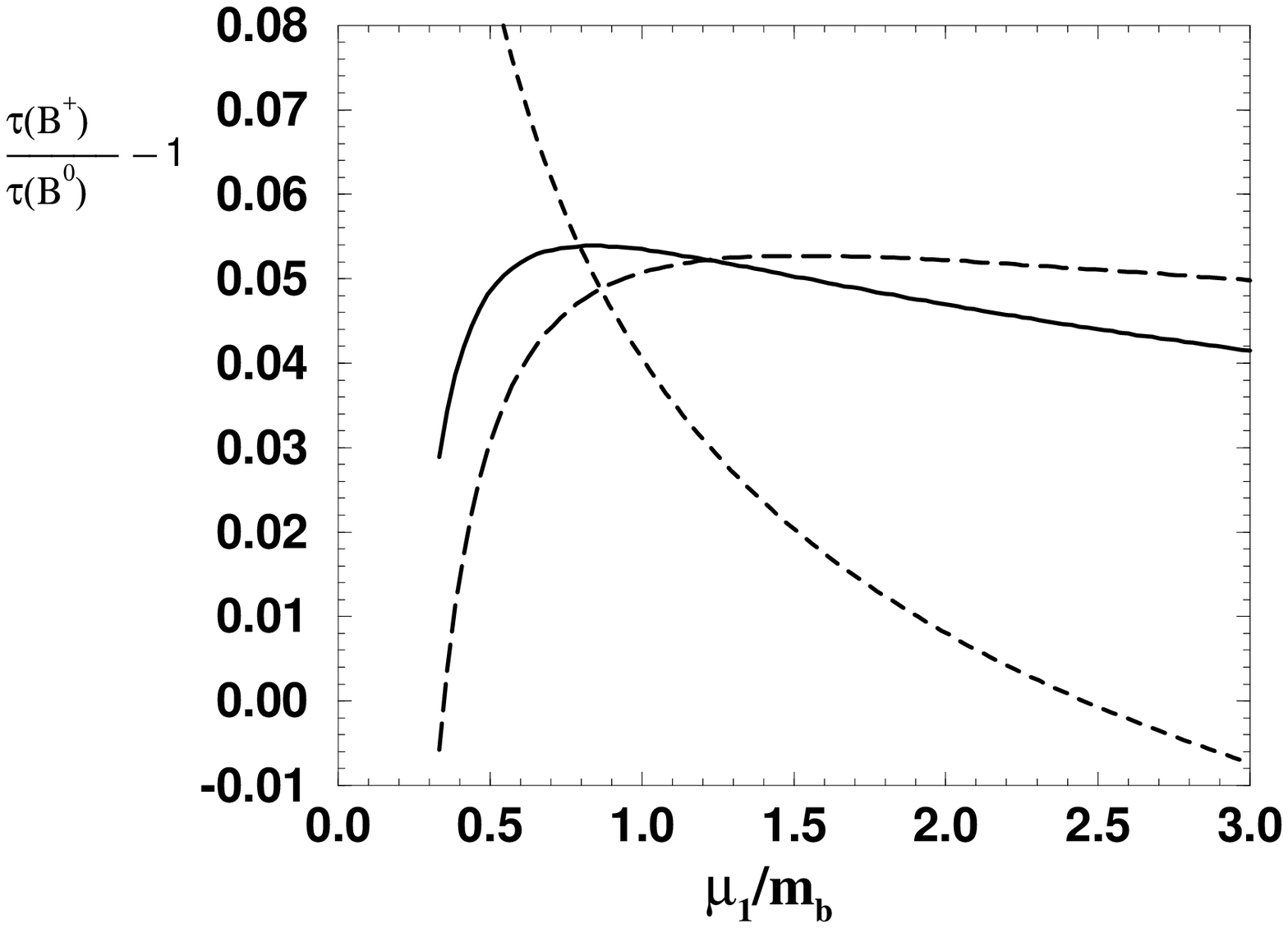}
}
\caption{Dependence of $\tau(B^+)/\tau(B^0_d)-1$
  on $\mu_1/m_b$ for the central values of the input parameters and
  $\mu_0=m_b$.  The solid (short-dashed) line shows the NLO (LO)
  result. The long-dashed line shows the NLO result in the
  approximation of \cite{rome}, i.e.\ $z$ is set to zero in the NLO
  corrections.  }\label{fig:ldplot}
\end{nfigure}

\boldmath
\subsection{$\tau(\Xi_b^0)/\tau(\Xi_b^-)$}\label{subs:xi}
\unboldmath%
The SU(3)$_{\rm F}$ anti-triplet $(\Lambda_b \sim bud,\, \Xi_b^0 \sim
bus,\, \Xi_b^- \sim bds)$ comprises the $b$-flavoured baryons whose
light degrees of freedom are in a $0^+$ state. These baryons decay
weakly.  Baryon lifetimes have attracted a lot of theoretical
attention: the measured $\Lambda_b$ lifetime falls short of
$\tau(B_d^0)$ by roughly 20\% \cite{pdg}, which has raised concerns
about the applicability of the HQE to baryons. Unfortunately this
interesting topic cannot yet be addressed at the NLO level, because
$\tau(\Lambda_b)/\tau(B_d^0)$ receives contributions from the
SU(3)$_{\rm F}$-singlet portion ${\cal T}_{sing}$ of the transition
operator in \eq{t3} and NLO corrections to ${\cal T}_{sing}$ are unknown at
present.  Further the hadronic matrix elements entering
$\tau(\Lambda_b)/\tau(B_d^0)$ involve penguin contractions of the
operators in \eq{ops}, which are difficult to compute. It is, however,
possible to predict the lifetime splitting within the iso-doublet
$(\Xi_b^0,\Xi_b^-)$ with NLO precision. The corresponding LO diagrams
are shown in Fig.~\ref{fig:bar}.
\begin{nfigure}{t}
\centerline{
\epsfxsize=0.8\textwidth
\epsffile{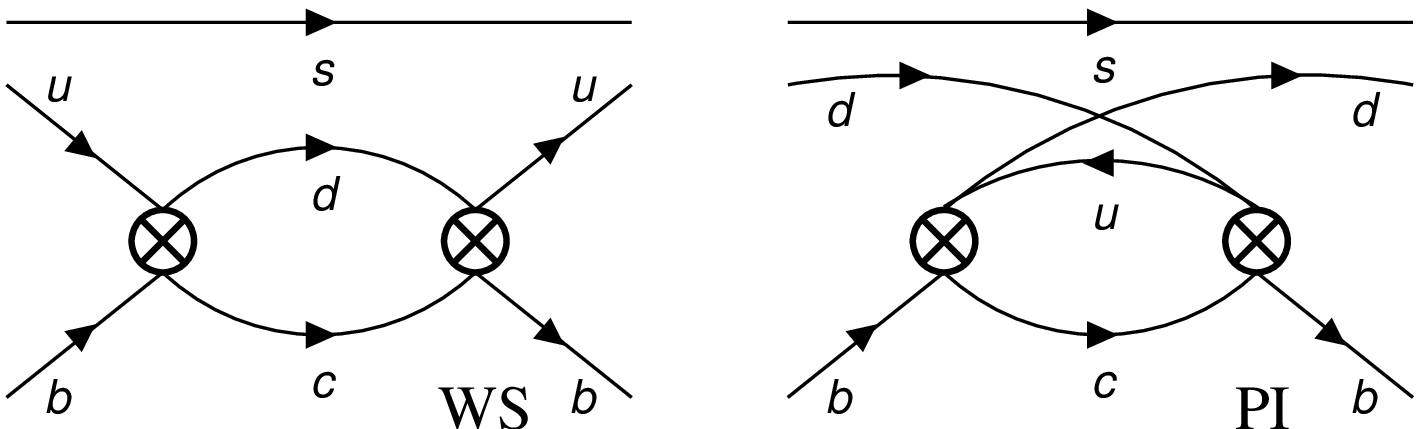}
}
\caption{\textit{Weak scattering}\ (WS) 
  and PI diagrams for $\Xi_b$ baryons in the leading order of QCD. They
  contribute to $\Gamma (\Xi_b^0)$ and $\Gamma (\Xi_b^-)$, respectively.
  CKM-suppressed contributions are not shown.  }\label{fig:bar}
\end{nfigure} 
For $\Xi_b$'s the weak decay of the valence $s$-quark could be
relevant: the decays $\Xi_b^- \to \Lambda_b \pi^-$, $\Xi_b^- \to
\Lambda_b e^- \ov{\nu}_e$ and $\Xi_b^0 \to \Lambda_b \pi^0$ are
triggered by $s\to u$ transitions and could affect the total rates at
the ${\cal O}(1\%)$ level \cite{v}. Once the lifetime measurements
reach this accuracy, one should correct for this effect. To this end
we define
\begin{eqnarray}
\ov{\Gamma} (\Xi_b) \! &\equiv& \! 
  \Gamma (\Xi_b) - \Gamma (\Xi_b \to \Lambda_b X)
  \; = \; \frac{1- B(\Xi_b \to \Lambda_b X)}{\tau(\Xi_b)} 
  \; \equiv \; \frac{1}{\ov{\tau}(\Xi_b)}
  \quad
  \mbox{for }\, \Xi_b=\Xi_b^0,\Xi_b^-, \qquad
\end{eqnarray}
where $B(\Xi_b \to \Lambda_b X)$ is the branching ratio of the
above-mentioned decay modes. Thus $\ov{\Gamma} (\Xi_b)$ is the
contribution from $b\to c$ transitions to the total decay rate.
In analogy to \eq{diffg} one finds
\begin{eqnarray}
\ov{\Gamma} (\Xi_b^-) - \ov{\Gamma} (\Xi_b^0) & = & 
  \frac{G_F^2 m_b^2 |V_{cb}|^2}{12 \pi } \, f_B^2 M_B \, 
        \lt( |V_{ud}|^2 \vec{F}{}^{\WA} + |V_{cd}|^2 \vec{F}{}^{c} 
         - \vec{F}{}^{\PI} \rt) \cdot \vec{B}{}^{\Xi_b} 
    . \label{diffgxi}
\end{eqnarray}
Here $\vec{B}{}^{\Xi_b} = (L_1^{\Xi_b}(\mu_0),L_{1S}^{\Xi_b}(\mu_0),
L_{2}^{\Xi_b}(\mu_0),L_{2S}^{\Xi_b}(\mu_0))^T$ comprises the hadronic
parameters defined as
\begin{eqnarray}
\langle \Xi_b^0 | (Q^u - Q^d) (\mu_0) | \Xi_b^0 \rangle 
& = & f_B^2 M_B M_{\Xi_b} \, L_1^{\Xi_b} (\mu_0), \nn
\langle \Xi_b^0 | (Q_S^u - Q_S^d) (\mu_0) | \Xi_b^0 \rangle 
& = & f_B^2 M_B M_{\Xi_b} \, L_{1S}^{\Xi_b} (\mu_0), \nn
\langle \Xi_b^0 | (T^u - T^d) (\mu_0) | \Xi_b^0 \rangle 
& = & f_B^2 M_B M_{\Xi_b} \, L_2^{\Xi_b} (\mu_0), \nn
\langle \Xi_b^0 | (T_S^u - T_S^d) (\mu_0) | \Xi_b^0 \rangle 
& = & f_B^2 M_B M_{\Xi_b} \, L_{2S}^{\Xi_b} (\mu_0) . \label{l} 
\end{eqnarray}
In contrast to the $B$ meson system, the four matrix elements in
\eq{l} are not independent at the considered order in
$\Lambda_{QCD}/m_b$. Since the light degrees of freedom are in a
spin-0 state, the matrix elements $\langle \Xi_b | 2Q_S^q+Q^q | \Xi_b
\rangle$ and $\langle \Xi_b | 2T_S^q+T^q | \Xi_b \rangle$ are
power-suppressed compared to those in \eq{l} (see e.g.\ 
\cite{hqe,ns}). This, however, is not true in all renormalization
schemes, in the $\ov{\rm MS}$ scheme used by us $2Q_S^q+Q^q$ and
$2T_S^q+T^q$ receive short-distance corrections, because hard gluons
can resolve the heavy $b$-quark mass.  This feature is discussed in
\cite{bbgln}. These short-distance corrections are calculated from the
diagrams $E_1\ldots E_4$ in Fig.~\ref{fig:nlo}. For our scheme we find
\begin{eqnarray}
\lt( \begin{array}{c}
        L_{1S}^{\Xi_b}  (m_b)\\
        L_{2S}^{\Xi_b}  (m_b)
      \end{array}
\rt)
&=& 
\lt[ - \frac{1}{2} \; + \;  \frac{\alpha_s(m_b)}{4 \pi} \, 
\lt( \begin{array}{cc}
        -28/3  & -7  \\
        -14/9  & 7/2  
      \end{array}
\rt) \rt] \, 
\lt( \begin{array}{c}
        L_1^{\Xi_b}  (m_b)\\
        L_2^{\Xi_b}  (m_b)
     \end{array}
\rt)  \; + \; 
      {\cal O} \Big( \frac{\Lambda_{QCD}}{m_b} \Big). 
      \quad \label{sdl}
\end{eqnarray}
As an important check we find that the dependence on the infrared regulator
drops out in \eq{sdl}.  With \eq{sdl} we can express the width
difference in \eq{diffgxi} in terms of just the two hadronic
parameters $L_1^{\Xi_b}$ and $L_2^{\Xi_b}$.  We find
\begin{eqnarray}
  \frac{\ov{\tau}(\Xi_b^0)}{\ov{\tau}(\Xi_b^-)} - 1 & = &
    \ov{\tau}(\Xi_b^0) \, 
                       \lt[ \Gamma (\Xi_b^-) - \Gamma (\Xi_b^0) \rt] \nn
& = & 0.59   \, \lt( \frac{|V_{cb}|}{0.04} \rt)^2  
        \, \lt( \frac{m_b}{4.8\gev} \rt)^2 \, 
           \lt( \frac{f_B}{200\mev} \rt)^2 \, 
      \frac{\ov{\tau}(\Xi_b^0)}{1.5\, {\rm ps}} 
        \times \nn 
&& \qquad \qquad \qquad
      \Big[ \, ( 0.04 \pm 0.01) \, L_1  \; - \; 
               ( 1.00 \pm 0.04) \, L_2 \, \Big] 
               ,~~~ \label{phentxi}
\end{eqnarray}
with $L_i=L_i^{\Xi_b}(\mu_0=m_b)$. For the baryon case there is no
reason to expect the color-octed matrix element to be much smaller
than the color-singlet ones, so that the term with $L_2$ will 
dominate the result. The hadronic parameters $L_{1,2}$ have been
analysed in an exploratory study of lattice HQET \cite{dsm} for
$\Lambda_b$ baryons. Up to SU(3)$_{\rm F}$ corrections, which are
irrelevant in view of the other uncertainties, $L_i^{\Xi_b}$ and
$L_i^{\Lambda_b}$ are equal.

\section{Conclusions}

We have computed the Wilson coefficients
in the heavy quark expansion to order $(\Lambda_{QCD}/m_b)^3$
for the $B^+$--$B_d^0$ lifetime difference at next-to-leading
order in perturbative QCD.
These coefficients depend on the scheme and scale $\mu_0$ used
to define the matrix elements of the $\Delta B=0$ operators in the
effective theory. Our scheme is specified by the NDR prescription
for $\gamma_5$, $\overline{\rm MS}$ subtraction and the definition
of evanescent operators given in (\ref{defev}).
The ${\cal O}(\alpha_s)$ accuracy is crucial for a satisfactory
matching of the Wilson coefficients to the matrix elements
determined with lattice QCD. Current lattice calculations,
which are still in a relatively early stage in this case,
yield, when combined with our calculations, 
$\tau(B^+)/\tau(B^0_d)=1.053\pm 0.016 \pm 0.017$ [see (\ref{res})]. 
The effects of 
unquenching and $1/m_b$ corrections are not yet included, 
but could well be on the order of 0.05. 
Next-to-leading order corrections to $\tau(B^+)/\tau(B^0_d)$ 
were recently computed in the approximation $m_c=0$ \cite{rome}. 
Taking the limit $m_c\to 0$ of our results we find agreement 
with this calculation. 

A substantial improvement of the NLO calculation is the large 
reduction of perturbative uncertainty reflected in the scale
dependence of $\Delta B=1$ Wilson coefficients from the
standard weak Hamiltonian. This scale dependence had been found to
be very large at leading order, preventing even an unambiguous
prediction of the sign of $\tau(B^+)/\tau(B^0_d)-1$ up to 
now \cite{ns}. 
With this major source of uncertainty removed by the NLO calculation,
further progress will depend on continuing advances in the
evaluation of the nonperturbative hadronic matrix elements and 
the computation of $1/m_b$-suppressed effects.

\end{document}